\documentclass[preprint,aps,twocolumn,10pt]{revtex4}
\usepackage{amsmath,amssymb,amsthm,graphicx,latexsym}
\newcommand{\doublespacing}{\let\CS=\@currsize\renewcommand{\baselinesstrech}
{2.0}\tiny\CS}

\newcommand{\bd}{\begin{document}}
\newcommand{\ed}{\end{document}}
\newcommand{\bc}{\begin{center}}
\newcommand{\ec}{\end{center}}
\newcommand{\vs}{\vspace}

\begin{document}

\title{\Large \bf Classical Trajectories of the Continuum States
of the ${\cal{PT}}$ symmetric Scarf II potential }

\vs{.3cm}

\author{{\large Anjana Sinha}$^*$ \\
Department of Applied Mathematics, Calcutta University, 92 A.P.C.
Road, Kolkata - 700 009, INDIA }

\vs{1cm}


\begin{abstract}

\noindent We apply the factorization technique developed by Kuru
et. al. [Ann. Phys. {\bf 323} (2008) 413] to obtain the exact
analytical classical trajectories and momenta of the continuum
states of the non Hermitian but ${\cal{PT}}$ symmetric Scarf II
potential. In particular, we observe that the strange behaviour of
the quantum version at the spectral singularity has an interesting
classical analogue.

\vs{.5cm}


\vs{.5cm}


\vs{.3cm}

\noindent $^*$ e-mail : anjana23@rediffmail.com ,
sinha.anjana@gmail.com

\end{abstract}


\maketitle

\newpage

\section{Introduction}

Non Hermitian Hamiltonians, especially those with ${\cal{PT}}$
symmetry or space time reflection symmetry, aroused curiosity
among the scientific community ever since Bender et. al. showed
that one can replace the condition of Dirac Hermiticity of the
Hamiltonian (viz., $H^{\dagger} = H$) by the ${\cal{PT}}$ symmetry
of the same (i.e. $H^{\cal{PT}} = H$), and yet obtain real and
discrete spectrum under certain conditions
\cite{bender-1,bender-2}. Since then scientists have extensively
studied such systems \cite{pt-homepage}, with an attempt to extend
the framework of conventional quantum mechanics into the complex
domain [4-11]. Theoretical predictions of such systems are found
in quantum field theory, mathematical, atomic and solid state
physics, classical optics, etc. \cite{pt-appl}, while ${\cal{PT}}$
symmetric optical lattices have provided the ground for
experimental verification [13-20].

The study of complex classical mechanics play an important role in
understanding the classical limit of complex quantum theories, as
it is generally observed that quantum mechanics and classical
mechanics provide profoundly different descriptions of the
physical world in many cases. In an attempt to bridge this gap,
several studies (both analytical and numerical) of the classical
trajectories $x(t)$ of complex Hamiltonians were taken up, and
many interesting observations were reported. For example, it was
found in some cases that the complex classical trajectories of a
particle having real energy are closed and periodic, while those
for a particle having complex energy are open and irregular
\cite{bender-rpp,zafar-class,dd-as-pr}. However, this does not
hold in general, as it was observed that for a special discrete
set of curves in the complex-energy plane, the classical orbits
are actually periodic, because of the possibility of some sort of
quantization condition \cite{bender-complex-per}. Additionally,
attempts have even been made to explain the concept of tunnelling
with the help of classical trajectories \cite{bender-tunn}.

However, all the previous studies of complex classical systems
have been for bound states. In one of our previous studies, we had
obtained the exact analytical classical trajectories of the bound
states of the ${\cal{PT}}$ symmetric Scarf II potential
\cite{dd-as-pr}. The main observation in ref. \cite{dd-as-pr} was
that the switching of energy values from real to complex ones at
the ${\cal{PT}}$ transition point in the quantum picture, was
accompanied by the changing of the corresponding classical
trajectories from closed periodic curves to open irregular ones.
Naturally, our next attempt is to obtain the exact analytical
classical trajectories of the scattering states of the same
potential. In particular, our main aim in the present work is to
see if spectral singularity in the quantum picture reveals any
interesting feature in the corresponding classical version. For
this purpose, we shall follow the factorization technique
developed by Kuru et. al. \cite{kuru-annals}. The motivation for
studying the Scarf II potential is varied --- \\
1. The problem is exactly solvable both quantum mechanically as
well as classically \cite{dd-as-pr}. \\
2. The bound state spectrum exhibits the interesting phenomenon of
${\cal{PT}}$ phase transition at the exceptional point, where
the real spectrum enters the complex domain \cite{zafar-scarf}. \\
3. The scattering state spectrum shows a spectral singularity at a
single positive energy, where both reflection and transmission
coefficients diverge and the eigen states are no longer linearly
independent \cite{zafar-ss,mostafazadeh-ss}. \\
4. This potential has been used to study nonlinear optical beam
dynamics in a single ${\cal{PT}}$ complex crystal
\cite{pt-optics2}.

\vspace{.3cm}

The organization of the paper is as follows. In Section II, we
briefly introduce the formalism applied in ref.
\cite{kuru-annals}, and extend the same to the complex domain. We
apply it to obtain the exact analytical classical trajectories for
the classical analogue of the real (Hermitian) Scarf II potential
in Section III, and for the ${\cal{PT}}$ symmetric potential in
Section IV. Section V is specially devoted to the classical
picture at the spectral singularity. Finally, Section VI is kept
for Conclusions and Discussions.

\section{Formalism}

This section is primarily included to make the work
self-contained. To briefly discuss the formalism applied in ref.
\cite{kuru-annals}, we start with the one dimensional classical
Hamiltonian (in units $ \hbar = 2m=1$)
\begin{equation}\label{h}
    H (x,p) = p ^2 + V(x)
\end{equation}
where $V(x)$ denotes the potential, and $x$ and $p$ are the
canonical coordinates of position and momentum, respectively, with
their Poisson bracket $ \displaystyle \{ x,p \} = 1$. Hence, the
classical particle obeys the equations of motions as given by the
Hamilton's equations
\begin{equation}\label{xp}
    \dot{x} = \displaystyle \frac{\partial H}{\partial p} = 2p
    \qquad , \qquad
    \dot{p} = \displaystyle - \frac{\partial H}{\partial x} =
    - V^{\prime}(x)
\end{equation} \\
so that $ \ \ \ddot{x} = 2 \dot{p} = 2 V^{\prime} (x) \ \ $ which
on integration gives the velocity of the particle as
\begin{equation}\label{v}
    \displaystyle v = \frac{dx}{dt} = \pm 2 \sqrt{ E - V(x)}
\end{equation}
$E$ being its energy. While extending the formalism to the complex
domain, the particle is expected to lie in the complex plain, so
that the path $x(t)$ it traces out, as well as its velocity
$v(t)$, may take complex values. The initial conditions determine
the initial velocity of the particle, and any point in the complex
plane may be taken as an initial starting point
\cite{bender-class}.

Following the factorization technique of ref. \cite{kuru-annals},
we assume a factorization of the Hamiltonian $H$ in the form
\begin{equation}\label{h-aa}
    H = A^+ A^- + \gamma (H)
\end{equation}
In usual quantum mechanical factorizations, $\gamma$ is the
factorization constant, but in eq. (\ref{h-aa}) above, $\gamma
(H)$ may depend on $H$. Furthermore, $A^{\pm}$ are no longer
complex conjugates. $A^{\pm}$ are taken to be of the form
\begin{equation}\label{aa}
    A^{\pm} = \mp i f(x) p + \sqrt{H} g(x) + \varphi (x) + \phi
    (H)
\end{equation}
so that $A^{\pm}$ and $H$ are assumed to define a deformed
algebra
\begin{equation}\label{ha-poisson}
    \displaystyle \left \{ A^{\pm} , H \right \} = \pm i \alpha
    (H) A^{\pm} \ , \
    \displaystyle \left \{ A^+ , A^- \right \} = - i \beta (H)
    A^{\pm}
\end{equation}
The unknown functions are determined by solving the following
equations simultaneously
\begin{equation}\label{f}
    \displaystyle f(x) = \displaystyle \frac{2}{\alpha (H)} \left[ \varphi
    ^{\prime} (x) + g^{\prime} (x) \sqrt{H} \right]
\end{equation}
\begin{equation}\label{f-prime}
\begin{array}{lll}
    & & \displaystyle f(x) V^{\prime} (x) - 2 f^{\prime} (x)
    \left[ H-V(x) \right] \\
    &=& \displaystyle \alpha (H) \left\{ g(x) \sqrt{H} +
    \varphi(x) + \phi (H) \right]
\end{array}
\end{equation}
\begin{equation}\label{beta}
\begin{array}{lll}
    \beta &=& \displaystyle 2 \sqrt{H} \left[ f^{\prime} (x) g(x)
    - f(x) g^{\prime} (x) \right] \\
    &-& \displaystyle \frac{1}{\sqrt{H}} g(x)
    \left[ 2 f^{\prime} (x) V(x) + f(x) V^{\prime} (x) \right] \\
    \\
    &+& \displaystyle \ 4 f^{\prime}(x) \phi ^{\prime} (H) \left[ H-V(x)
    \right] \\
    &-& \displaystyle  2 f(x) \left[ \varphi ^{\prime} (x) + \phi ^{\prime} (H)
    V^{\prime} (x) \right]
\end{array}
\end{equation}
Now we construct two time dependent integrals of motion of the
form
\begin{equation}\label{qpm}
    Q^{\pm} = \displaystyle A^{\pm} e^{\mp i \alpha(H) t}
\end{equation}
so that the values of $Q^{\pm}$ may be denoted by
\begin{equation}\label{q}
    q^{\pm} = \displaystyle c(E) e^{\pm i \theta _0}
\end{equation}
where $\theta _0$ is determined from initial conditions, and
\begin{equation}\label{c}
    c(E) = \displaystyle \mid q^{\pm} \mid = \sqrt{E-\gamma (H) }
\end{equation}
For $c(E)$ to be real, the expression within the square root sign
must be positive. This condition gives the range of energy values
for the classical particle. It is worth noting here that for bound
and scattering states, the complex character of the factors
$A^{\pm}$ may change and the deformed algebra may also be
different \cite{kuru-annals}.


\section{Classical analogue of the Real Scarf II potential}

This section is included to provide a direct comparison between
the classical pictures of the corresponding Hermitian and
${\cal{PT}}$ symmetric Scarf II potentials. For bound states ({\it
i.e.}, $E < 0 $), $A^{\pm}$ are defined as
\begin{equation}\label{aa-unbound}
    A^{\pm} = \mp i f(x) p + \sqrt{-H} g(x) + \varphi (x) + \phi
    (-H)
\end{equation}
For unbounded motion ($E > 0$), putting $ \sqrt{-E} = i \sqrt{\mid
E \mid} $, changes the algebra of the Poisson bracket to
\begin{equation}\label{a0-apm}
    \displaystyle \left \{ A^{\pm} , A_0 \right \}
    = \displaystyle \pm i \frac{\alpha _0}{2} A^{\pm} \ , \
    \displaystyle \left \{ A^+ , A^- \right \} = i \alpha _0
    A_0
\end{equation}
where $ \displaystyle A_0 = - i \sqrt{H}$.

\noindent The final expression for $V(x) $ should be of the form
\begin{equation}\label{v-scarf}
    V(x) = \displaystyle  \gamma _0 \ {\rm{sech}}^2 \ {\frac{\alpha
    _0 x}{2}} + 2 \delta \   {\rm{sech}} \ {\frac{\alpha
    _0 x}{2}} \ \tanh {\frac{\alpha _0 x}{2}}
\end{equation}
We have intentionally considered the coefficient of $
{\rm{sech}}^2 \ \displaystyle {\frac{\alpha _0 x}{2}} $, viz., $
\gamma _0 $, to be positive, since the quantum mechanical
${\cal{PT}}$ symmetric version of eq. (\ref{v-scarf}) displays the
interesting phenomenon of spectral singularity. We shall return to
this point later, in Sections IV and V.  Eq. (\ref{v-scarf}),
demands that in the expression for $A^{\pm}$ in (\ref{aa-unbound})
we take the following forms of the functions $g(x) \ , \ \varphi
(x) \ , \ \phi (H) $ :
\begin{equation}\label{g-phi}
    g(x) \neq 0 \ , \ \varphi (x) = 0 \ , \
    \phi (H) = \displaystyle - i \frac{\delta}{\sqrt{H}}
\end{equation}
Solving equations (\ref{f}), (\ref{f-prime}) and (\ref{beta})
simultaneously, one of the possible options could be
\begin{equation}\label{gf-scarf}
    g(x) = \displaystyle \sinh {\frac{\alpha _0 x}{2}} \ , \
    f(x) = \displaystyle \cosh {\frac{\alpha_0 x}{2}}
\end{equation}
with
\begin{equation}\label{alpha-scarf}
    \alpha (H) = i \alpha _0 \sqrt{H}
\end{equation}
Putting
\begin{equation}\label{gamma}
    \gamma (H) = \displaystyle \gamma _0 + \frac{\delta ^2}{H}
\end{equation}
in eq. (\ref{h-aa}), we obtain after some simplification, the
Hermitian Scarf II potential
\begin{equation}\label{H-scarf2}
    \displaystyle H
    = p^2 + \gamma _0 \ {\rm{sech}} ^2 \frac{\alpha _0 x}{2} + 2  \delta
    \displaystyle \ {\rm{sech}} \frac{\alpha _0 x}{2} \ \tanh \frac{\alpha _0 x}{2}
\end{equation}
Hence
\begin{equation}\label{aa-mod}
     c(E) = \displaystyle \sqrt{ E - \gamma _0 -
     \frac{\delta  ^2}{E}}
\end{equation}
Since we are dealing with scattering states, $E$ is always
positive; therefore
\begin{equation}\label{e-real-range}
    E \ > \ \displaystyle \frac{\sqrt{\gamma _0 ^2 + 4 \delta
    ^2} + \gamma _0}{2}
\end{equation}
The integrals of motion $Q^{\pm}$, are pure imaginary; so values
of $Q^{\pm}$ take the form
\begin{equation}\label{q-value}
    q^{\pm} = \displaystyle \mp i \ c(E) \exp [\mp \theta _0]
\end{equation}
Straightforward calculations give the classical trajectories and
momenta for the classical analogue of the quantum mechanical
Hermitian Scarf II model as
\begin{equation}\label{x-real}
    x (t) = \displaystyle \frac{2}{\alpha _0} \sinh ^{-1} \left\{
    \frac{c(E)}{\sqrt{E}} \sinh \left( \theta _0 + \alpha _0
    \sqrt{E} t \right) + \frac{\delta}{E} \right\}
\end{equation}
\begin{equation}\label{p-real}
    p (t) = \displaystyle \frac{1}{\displaystyle \cosh \frac{\alpha _0 x}{2} }
        \ c(E) \ \cosh \left( \theta _0 + \alpha _0
    \sqrt{E} t \right)
\end{equation}
where $\theta _0$ denotes the initial starting point.

\noindent In Fig. 1, we plot the Hermitian Scarf II potential as
in eq. (\ref{v-scarf}), and in Fig. 2, we plot the  phase space
trajectories for this potential, for different positive energies.
Parameter values used are $ \alpha _0 = 2 , \ \delta = 2 , \
\gamma _0 = 6 $. The plot shows that as the particle comes near
the barrier, its classical momentum, and hence kinetic energy,
decreases, as expected.

{\begin{figure}[hp]
\begin{center}
\scalebox{0.6}{\includegraphics{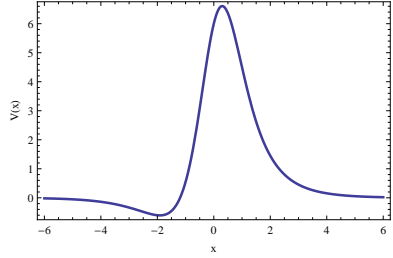}}
\label*{}\caption{\small {Plot of $V(x)$ against $x$ for Real
Scarf II pot.}}
\end{center}
\end{figure}}

{\begin{figure}[hp]
\begin{center}
\scalebox{0.4}{\includegraphics{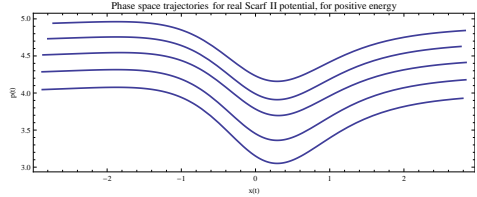}}
\label*{}\caption{\small {Plot of $p(t)$ against $x(t)$ for
classical analogue of Real Scarf II potential}}
\end{center}
\end{figure}}

\section{Classical analogue of the ${\cal{PT}}$ symmetric Scarf II potential}

For $\cal{PT}$ symmetric Scarf II potential, we put $ \delta = i
\delta_I $, so that
\begin{equation}\label{H-PT-scarf2}
    \displaystyle H
    = p^2  + \gamma _0 \ {\rm{sech}} ^2 \frac{\alpha _0 x}{2} +  i
    \ 2 \ \delta_I
    \displaystyle \ {\rm{sech}} \frac{\alpha _0 x}{2} \ \tanh \frac{\alpha _0 x}{2}
\end{equation}
For scattering states, this particular quantum mechanical model
exhibits the interesting phenomenon of spectral singularity at a
single positive energy $E_{s}$
\begin{equation}\label{e-s}
    E_{s} = \displaystyle \frac{1}{4} \left[ \mid 2 \delta _I  \mid
     - \left( \gamma _0 + \frac{1}{4} \right) \right]
\end{equation}
Spectral singularity (ss) is also known as zero-width resonance,
as both the transmission and reflection coefficients diverge at
this particular energy. Thus these (ss) may be seen as positive
energy discrete poles of transmission and reflection coefficients.
For the potential given above in eq. (\ref{H-PT-scarf2}), in the
quantum picture, ss occurs when the parameters $\gamma _0$ and
$\delta _I$ satisfy the following conditions
\cite{zafar-ss,mostafazadeh-ss}
\begin{equation}\label{eq-ss}
\begin{array}{lcl}
    \displaystyle \mid 2 \delta _I  \mid \ &>& \
    \gamma _0  \ + \ \displaystyle \frac{{\rm{sign \ of \ }}
    \delta _I}{4} \\ \\
    \displaystyle \gamma _0 + \mid 2 \delta _I  \mid
    &=& 4 n^2 + 4n + \displaystyle \frac{3}{4} \ , \ n = 0, 1, 2, \cdots \\
\end{array}
\end{equation}
The main motivation of the present study is to see if this feature
is manifested in the corresponding classical picture in any way.

\noindent For scattering states and hence positive energy, with
\begin{equation}\label{c-pt}
    c(E) = \displaystyle \sqrt{E - \gamma _0 + \frac{\delta _I ^2}{E} }
\end{equation}
the range of energy values is obtained as
\begin{equation}\label{e-range1-pt}
    {\rm{either}}  \ \qquad 0 \ < \ E \ < \ \displaystyle \frac{\gamma _0 -
    \sqrt{\gamma _0 ^2 - 4 \delta _I ^2}}{2}
\end{equation}
\begin{equation}\label{e-range2-pt}
     {\rm{or}} \ \ \qquad E \ > \ \displaystyle \frac{\gamma _0 +
    \sqrt{\gamma _0 ^2 - 4 \delta _I ^2}}{2}
\end{equation}
If the real part of the potential is greater than the imaginary
part, i.e., $ \gamma _0 \ > \ \mid 2 \delta _I \mid $, then energy
is always real. But in the reverse case, i.e., when the imaginary
part of the potential is greater the real part ($ \gamma _0 \ < \
\mid 2 \delta _I \mid $) then energy becomes complex, with a
positive real part. The interesting point to note here is that
even beyond the ${\cal{PT}}$ threshold (or phase transition
point), nonlinear states can still be found in the quantum version
of the Scarf II potential, with real eigenvalues
\cite{pt-optics2}.

{\begin{figure}[hp]
\begin{center}
\scalebox{0.7}{\includegraphics{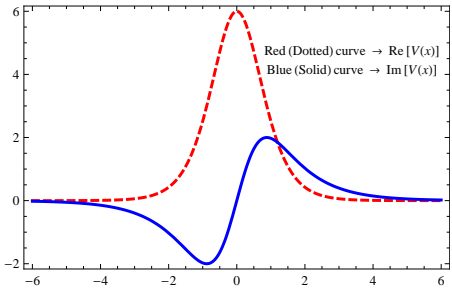}}
\label*{}\caption{\small {Plot of  $Re \ V(x)$ and $Im \ V(x) $
against $x$ for the ${\cal{PT}}$ symmetric Scarf II potential }}
\end{center}
\end{figure}}


\noindent Proceeding in a straightforward way, we obtain
expressions for the classical trajectories and momenta, similar to
those in equations (\ref{x-real}) and (\ref{p-real}), with $
\delta = i \delta _I  \ , \ c(E) = Re \ c(E) + i \ Im \ c(E) $ \\
    $  E = E_R + i E_I \ ,  \ x(t) = Re \ x(t) +
    i \ Im \ x(t) $ \\
    $ p(t) = Re \ p(t) + i \  Im \ p(t) $  , etc.
In Fig. 3, we plot the potential profile of the ${\cal{PT}}$
symmetric Scarf II Hamiltonian, for parameter values $ \alpha _0 =
2 , \ \delta _I = 2 , \ \gamma _0 = 6 $. In Figures 4, 5, 6 and 7,
we plot a series of classical trajectories and momenta for this
particular model, for both real and complex energy, taking the
energy to lie within the permissible range, as given in equations
(\ref{e-range1-pt}) and (\ref{e-range2-pt}).


{\begin{figure}[hp]
\begin{center}
\scalebox{0.5}{\includegraphics{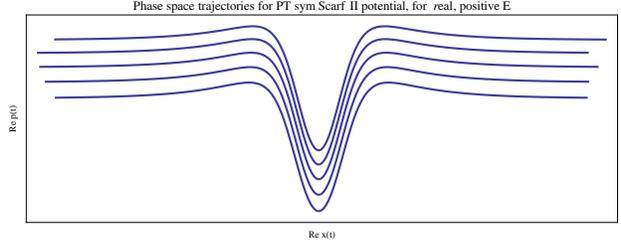}}
\label*{}\caption{\small {Plot of $Re \ p(t)$ against $Re \ x(t)$
for classical analogue of ${\cal{PT}}$ symmetric Scarf II
potential, for real energy, when $\gamma _0 \
> \ \mid 2 \delta _I \mid$ }}
\end{center}
\end{figure}}

\noindent Fig. 4 shows the real part of the phase space
trajectories for the classical analogue of the ${\cal{PT}}$
symmetric Scarf II potential, for same parameter values as in Fig.
3 : $ \alpha _0 = 2 , \ \delta _I = 2 , \ \gamma _0 = 6 $.
Analogous to the Hermitian case shown in Fig. 2, the particle
slows down at the potential barrier. In both Fig. 2 and Fig. 4,
the phase space curves are for different energies.

{\begin{figure}[hp]
\begin{center}
\scalebox{0.5}{\includegraphics{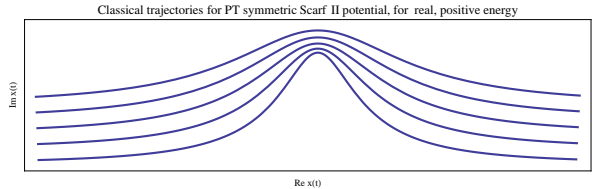}}
\label*{}\caption{\small {Plot of ${\rm{Im}} \ x(t)$ against $
{\rm{Re}} \ x(t)$ for classical analogue of ${\cal{PT}}$ sym.
Scarf II potential, for real energy}}
\end{center}
\end{figure}}

\noindent  Figures  5 and 6 show the classical trajectories traced
out by the particle, for real and complex energy respectively. The
parameter values are $ \alpha _0 = 2 , \ \delta _I = 2 , \ \gamma
_0 = 6 $ $(\gamma _0 \ > \ \mid 2 \delta _I \mid)$, for the
former, and $ \alpha _0 = 2 , \ \delta _I = 2 , \ \gamma _0 = 3 $
$(\gamma _0 \ < \ \mid 2 \delta _I \mid)$ for the latter. The
curves in Fig. 6 show a definite distortion from those in Fig. 5.
Evidentally, figures 5 and 6 have no counterpart in the classical
analogue of the Hermitian Scarf II model. Similar to the bound
state curves of ref. \cite{dd-as-pr}, the initial starting point
determines the trajectory, for the same energy.

\pagebreak

{\begin{figure}[hp]
\begin{center}
\scalebox{0.5}{\includegraphics{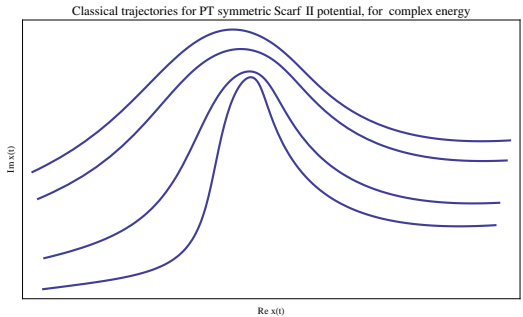}}
\label*{}\caption{\small {Plot of ${\rm{Im}} \ x(t)$ against $
{\rm{Re}} \ x(t)$ for classical analogue of ${\cal{PT}}$ sym.
Scarf II potential, for complex energy}}
\end{center}
\end{figure}}

\section{Classical picture at Spectral Singularity}

It may be mentioned that {\it spectral singularities} (ss) are
typical for complex scattering potentials, with no analogue in
Hermitian systems. A spectral singularity of $H$ occurs at a
positive energy $ E_{s} = k_{s} ^2  $ in the continuous spectrum
of $H$ where the Jost solutions $ \psi _{k \pm} (x)$, satisfying $
\psi _{k \pm} (x) \rightarrow \displaystyle e^{\pm i k x} $ for $
x \rightarrow \pm \infty  \ , \ E = k^2 $, become linearly
dependent and their Wronskian goes to zero \cite{mostafazadeh-ss}.
Thus, physically, spectral singularities correspond to scattering
states that behave like zero-width resonances. Spectral
singularities may be seen as positive energy discrete poles of
transmission and reflection coefficients. Physical realization of
spectral singularities are possible in ${\cal{PT}}$ symmetric
optical waveguides. If the frequency $\omega$ of the incoming wave
in such a waveguide can be tuned to the frequency $\omega _{s}$ of
the spectral singularity, then the amplitude of the outgoing wave
will be considerably enhanced as $ \omega \rightarrow \omega
_{s}$.

To study the phenomenon of spectral singularity in the classical
framework, we plot the real part of the phase space trajectories
for the ${\cal{PT}}$ symmetric Scarf II potential, in Fig. 7, for
different values of real energy, but when the real part of the
potential is smaller than the imaginary part, i.e.,
    $ \displaystyle \gamma _0 \ < \ \mid 2 \delta _I \mid $.
To be more precise, the plots are for parameter values $ \delta _I
= 12 , \ \gamma _0 = 4, \ \alpha _0 = 2$, but for different
positive energies. For this particular set of parameter values, it
is observed that for low energies (as shown in the inset figures
on below left), the classical particle is unable to overcome the
barrier, and the phase space trajectories show a discontinuity
near the potential barrier. As the energy increases the two
disjoint curves come closer together, showing the inaccessible
region for the classical particle decreases. Suddenly, as the
energy reaches a specific value viz., $E_s$ ($13.7$ for this
particular set of parameter values), the real part of the
classical momentum at the potential barrier tends to diverge.
Beyond this energy, the momentum decreases with increase in
energy, finally reaching a constant value. As compared to the
phase space trajectories in Fig. 4 ($\gamma _0 \ > \ \mid 2 \delta
_I \mid$), these plots are inverted --- the classical momentum
decreases as the particle moves away from the potential barrier.
We can identify this specific energy $E_s$ as the energy at which
classical spectral singularity takes place.

Thus the anomalous behaviour of a particle at the spectral
singularity is found in both the quantum and classical pictures,
though the manifestation is different in the two scenarios ---
while the reflection and transmission amplitudes blow up in the
quantum picture, the real part of the momentum tends to diverge in
the corresponding classical picture. This is the most important
finding of the present study. Another point worth noticing here is
that the condition for spectral singularity is slightly different
in the quantum and classical pictures. While the quantum condition
for ss is $ \mid 2 \delta _I \mid \ > \ \gamma _0 + \displaystyle
\frac{1}{4} $, the classical condition for the same turns out to
be $ \mid 2 \delta _I \mid \ > \ \gamma _0 $. This finding is
similar to the condition for exceptional point in the two pictures
\cite{dd-as-pr}.

{\begin{figure}[hp]
\begin{center}
\scalebox{0.8}{\includegraphics{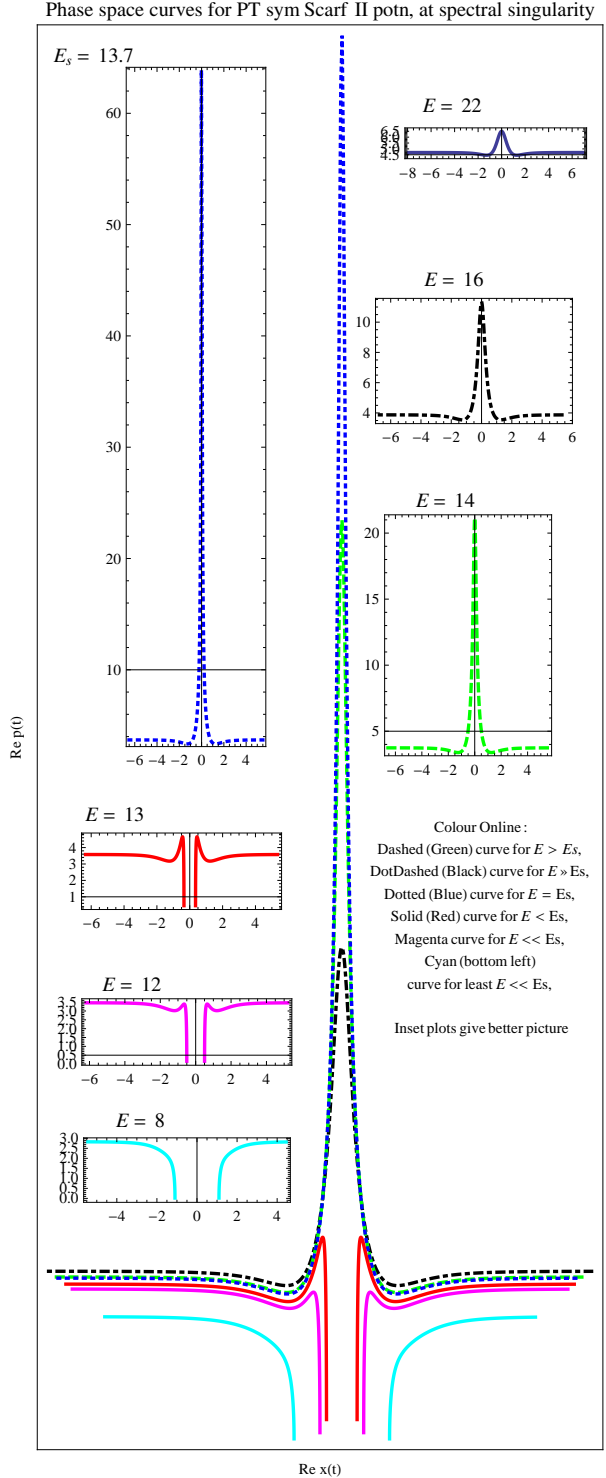}}
\label*{}\caption{\small {Colour Online : Plot of $Re \ p(t)$
against $ Re \ x(t)$ for classical analogue of ${\cal{PT}}$
symmetric Scarf II potential, exhibiting spectral singularity at a
single real energy $E_s=13.7$, when $\gamma _0 \ < \ \mid 2 \delta
_I \mid$ ; Inset curves are shown for a better understanding of
this phenomenon --- the 3 plots on below left are for $ E < E_s$
with the bottom most curve of least $E$, while the 3 on upper
right are for $E > E_s$ with the topmost curve of highest $E$ }}
\end{center}
\end{figure}}

\section{Conclusions}

To conclude, we have studied the scattering states (having
positive energy) of the classical analogue of an exactly solvable,
non Hermitian quantum mechanical Hamiltonian, viz., the
${\cal{PT}}$ symmetric Scarf II potential
    $$ V(x) = \displaystyle  \gamma _0 \ {\rm{sech}}^2 x - 2 i
    \delta _I \ {\rm{sech}} \ x \ \tanh x $$
by applying the factorization technique of ref.
\cite{kuru-annals}. In particular, we have obtained the exact
analytical classical trajectories of a particle subject to this
potential, and moving about in the complex $x$ plane. The main
motivation behind this study was to see if the interesting
phenomenon of spectral singularity exhibited in the quantum
version at a particular energy (say $E_s$), is manifested in the
classical picture as well.

We plotted a series of curves, for different parameter values, for
a clear picture of our findings. Figures 1 and 3 show the
Hermitian and ${\cal{PT}}$ symmetric Scarf II potentials,
respectively. Fig. 2 gives the classical phase space trajectories
for the former, while Fig. 4 gives the real part of the same for
the latter. That the two figures resemble each other is proof of
the fact that the complex ${\cal{PT}}$ symmetric potential shares
the classical features of the real (Hermitian) potential. Fig. 5
and Fig. 6 show the classical trajectories traced out by the
particle ($ Im \ x(t)$ against $ Re \ x(t)$), in the ${\cal{PT}}$
symmetric potential, for real and complex energy, respectively.
The point to be noted here is that none of the trajectories cross
each other in any plot.

\vs{.2cm}

Fig. 7 shows in detail the strange behaviour of the classical
particle at the spectral singularity, when the imaginary part of
the potential is greater than the real part, but energy is real.
So long as the energy of the classical particle is sufficiently
low, in fact lower than $E_s$, it is unable to cross the barrier,
and the phase space trajectories show a discontinuity near the
potential barrier. All of a sudden, at the specific energy for
spectral singularity $E_s$, the real part of the classical
momentum tends to diverge as the particle reaches the barrier.
Thereafter, the particle momentum at the barrier decreases with
increasing energy, thus showing a peculiar behaviour.

\vs{.2cm}

To give more stress to the strange behaviour of the classical
momentum near and at the spectral singularity (ss), and to see how
one obtains the specific energy ($E_s$) at ss, we plot the real
part of the classical momentum against energy, near the centre of
the potential barrier, in Fig. 8, for the same set of parameter
values as in Fig. 7 : $ \delta _I = 12 , \ \gamma _0 = 4, \ \alpha
_0 = 2$. For this particular set of parameter values, the spectral
singularity is found to occur in the classical picture at $E_s =
13.7$ . As the energy approaches $E_s$, the momentum rises
sharply, then falls rapidly with increasing energy, finally
attaining a constant value for large energy.

{\begin{figure}[hp]
\begin{center}
\scalebox{0.8}{\includegraphics{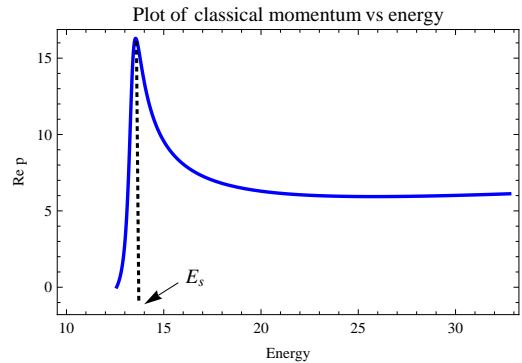}}
\label*{}\caption{\small {Plot of $Re \ p$ against $ Energy $ for
classical analogue of ${\cal{PT}}$ symmetric Scarf II potential,
exhibiting spectral singularity at a single real energy
$E_s=13.7$, when $ \delta _I = 12 , \ \gamma _0 = 4, \ \alpha _0 =
2$, i.e., $\gamma _0 \ < \ \mid 2 \delta _I \mid$ }}
\end{center}
\end{figure}}

In our earlier work on bound states of the ${\cal{PT}}$ symmetric
Scarf II potential \cite{dd-as-pr}, we had observed that at the
${\cal{PT}}$ transition point, also known as the exceptional
point, when the energy spectrum in the quantum picture goes from
the real to the complex domain, the classical trajectories switch
from closed, periodic orbits to open, irregular ones. In the
present study on the scattering states for the same potential, we
observe that at the spectral singularity, when reflection and
transmission coefficients blow up in the quantum version, the
classical momentum tends to diverge. Additionally, similar to the
condition for exceptional point \cite{dd-as-pr}, the classical
condition for spectral singularity differs from the corresponding
quantum condition by a factor of $\displaystyle \frac{1}{4}$.

Our next attempt would be to study more such cases, to check
whether this strange behaviour of the classical particle at the
spectral singularity is characteristic of this particular
potential, or some generalized conclusion can be drawn regarding
the classical behaviour at the spectral singularity.

\section{Acknowledgement}  Financial support for this work
was provided for by the Department of Science and Technology,
Govt. of India, through its grant SR/WOS-A/PS-06/2008. The author
thanks Prof. P. Roy for some interesting discussions, and the
unknown referee for some useful comments and suggestions.

\end{document}